\UseRawInputEncoding
\documentclass[journal]{IEEEtran}
\usepackage{xcolor,soul,framed} %,caption
\usepackage{graphicx}
\usepackage{amsmath}
\usepackage{array}
\usepackage{mdwmath}
\usepackage{mdwtab}
\usepackage{eqparbox}
\usepackage{url}
\usepackage{lineno,hyperref}
\usepackage{amsmath,amssymb}
\usepackage{booktabs}
\usepackage{makecell}
\usepackage{multirow}
\usepackage{multirow}
\usepackage{tabu,bm}
\usepackage{color}
\usepackage{lipsum,cuted}
\usepackage{stfloats}
\usepackage{cite}
\usepackage{enumitem}
\usepackage{diagbox}
\usepackage{subcaption}
\title{A Generalized Semantic Communication System: from Sources to Channels}

\author{Zhijin Qin, Feifei Gao, Bo Lin, Xiaoming Tao, Guangyi Liu, and  Chengkang Pan
\thanks{Zhijin Qin and Xiaoming Tao (corresponding author) are with the Department of Electronic Engineering, Tsinghua University£¬ Beijing, China. (e-mail: qinzhijin@tsinghua.edu.cn;taoxm@tsinghua.edu.cn) }
\thanks{Feifei Gao and Bo Lin are with the Department of Department of Automation, Tsinghua University, Tsinghua University, Beijing, China. (e-mail: feifeigao@ieee.org;linb20@mails.tsinghua.edu.cn) }
\thanks{Guangyi Liu and Chengkang Pan  are with China Mobile Research Institute, Beijing, China. (e-mail: liuguangyi@chinamobile.com; panchengkang@chinamobile.com) }
}

\begin{document}

\maketitle
\begin{abstract}
Semantic communication is regarded as the breakthrough beyond the Shannon paradigm, which transmits  the semantic information only to  improve the communication efficiency significantly. This article first introduces a framework for the generalized semantic communication system, which exploits the semantic information in both the multimodal source  and the wireless channel environment. Subsequently, the  deep learning enabled end-to-end semantic communications  and the environment semantics aided wireless communications  are demonstrated through two user cases. The article is concluded with several research challenges to boost the development of such a generalized semantic communication system.
\end{abstract}

\section{Introduction}
The concept of semantic communications could be traced to 1940s, when Shannon and Weaver  categorised communications into three levels~\cite{shannon1959mathematical}. The typical communication system is designed for the level one communication to address the successful transmission of symbols with the bit-error rate (BER) or symbol-error rate (SER) as  performance metrics. Semantic communication, categorized as the level two communication, focuses on the successful transmission of semantic meaning. Ever since the concept of semantic communication was proposed,  various paths have been investigated. However, a well-defined semantic communication theory is yet to develop.

Inspired by the boom of intelligent communications,  deep learning (DL) has shown its overwhelming privilege in semantic representation, compression, and  transmission~\cite{qin2021semantic}, which relaxes the requirements on general mathematical models for constructing the semantic theory. Different from typical communications, semantic communications introduce a new domain, i.e., semantic domain, to process and transmit information, in which only the essential information useful for serving intelligent tasks at the receiver is transmitted. By doing so, the size of the transmit data will be reduced significantly and the network could be tailed for serving different tasks at the receiver. Such characteristics of semantic communications make them naturally fulfill the requirements of future networks in terms of coordination, intelligence, and personalizing, which shows the great potential for supporting various applications, such as video conferencing, mixed reality (MR), and  meta-verse.

Most existing semantic communications focus on the semantic information processing conveyed in the source by treating wireless channels the same as that in typical communications. However,  the semantic information conveyed in wireless channel environment could also be utilized to reduce the cost of acquiring channel state information (CSI), which is essential to implement the communication system in a highly efficient manner. Due to the increasing demand for data transmission, the number of antennas proliferates, and thus the size of the channel matrix becomes larger. Hence, the pilot sequences required for channel estimation become longer, which consumes more spectrum resource and increase the latency of the transmission. Channel semantics opens a new dimension to acquire CSI and the corresponding study is on timely demand.

In this article, we propose a generalized semantic communications to exploit semantics in both sources and channels, which takes a different view from  existing works on semantic communications. The rest of this article is organized as follows. We first introduce the  generalized semantic communication  framework in Section II. To support such a framework, the DL-enabled end-to-end semantic communication techniques for multimodal source  are presented in Section III, while the environment semantics aided communication systems are detailed in Section IV. The final section concludes this article with several identified research challenges.

\section{A Generalized Semantic Communications}
For a simple wireless communication system, the received signal can be expressed as
\begin{equation}\label{HS}
   Y = Hx + n,
\end{equation}
where $H$ represents the channel gain, $x$ is the transmission signal, and $n$ is the channel noise. In the past several decades, we have witnessed the boom of wireless communications and their applications. Great efforts have been made to address the challenges in estimating $H$ with lower costs and recovering $x$.

\begin{figure}[t]
    \centering
    \includegraphics[width=9cm]{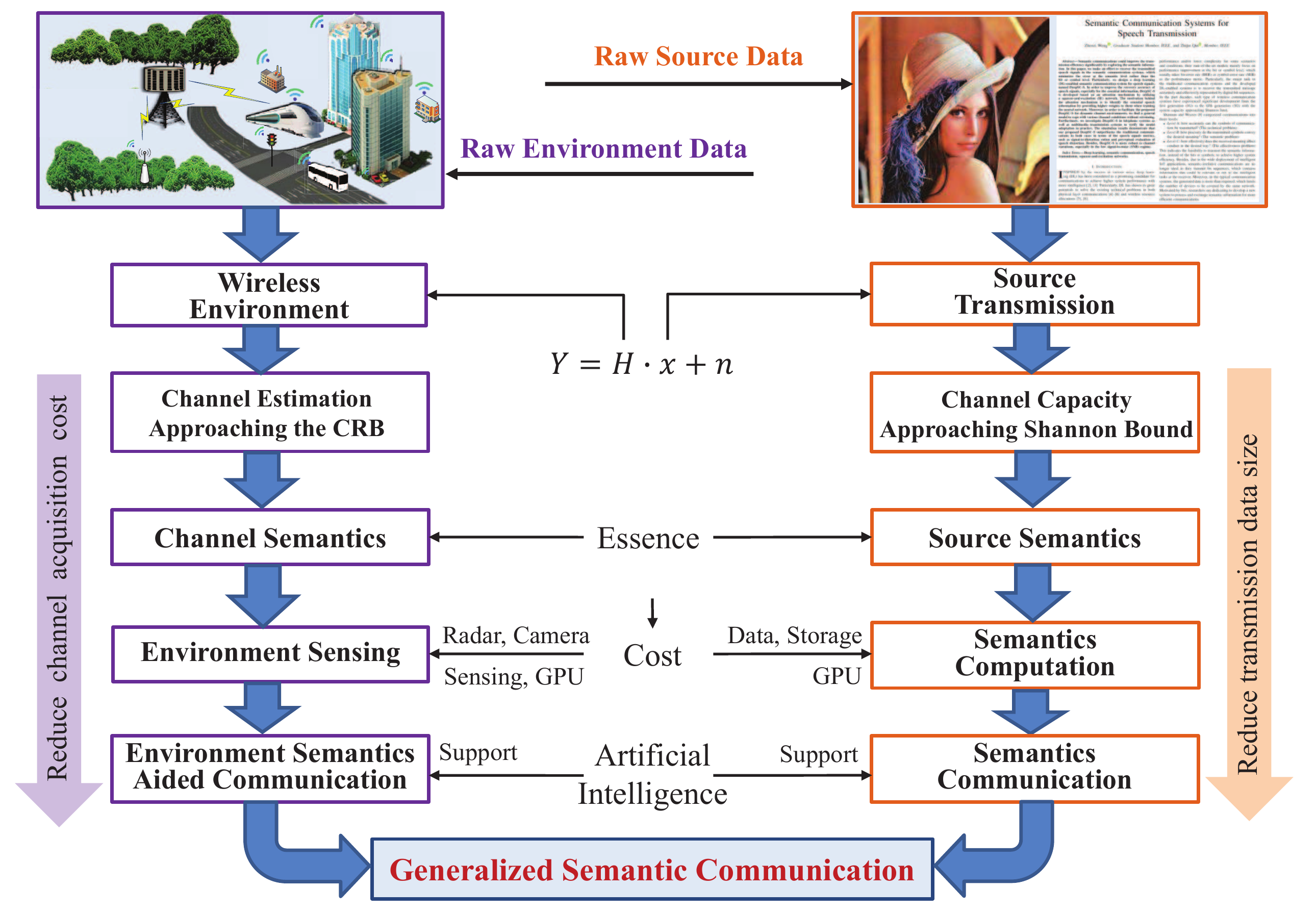}
    \caption{The proposed framework for generalized semantic communications.}
    \label{fig1}
\end{figure}

As shown in Fig. \ref{fig1}, for the source transmission, various techniques, such as source coding, channel coding, and modulation have been developed. Particularly, Shannon channel capacity provides the achievable upper bound for a point-to-point communication system. Thanks to the dedicated efforts made by researchers in the past several decades, the current communication system is approaching to its limit. As we can easily see from Shannon theorem,  typical methods to  improve the transmission  rate include increasing transmit power, adopting wider bandwidth, e.g., using millimeter wave (mmWave)  or even terahertz (THz) frequency, and introducing more antennas. However, the existing communication system still faces some critical challenges, such as spectrum shortage and heavy power consumption. It is desired to explore the semantic domain, in addition to the time, space, and frequency domain, to significantly improve communication efficiency.

Though the concept of semantic communication has been proposed long ago, its development has been rather limited in the past decades due to the lack of mathematical tools to handle semantic processing. With the aid of artificial intelligence (AI) techniques and high performance computing devices, semantic communication systems have been developed recently by using a deep neural network (DNN) to represent the transmitter and the receiver, respectively.

On the other side, the channel, another very important component in (\ref{HS}),  is characterised by the distribution of scatters and the electromagnetic propagation paths. Specifically, the parameters of each electromagnetic propagation path between the base station (BS) and the user are generated according to Maxwell's equations. Acquiring a precise channel is the prerequisite for high performance signal transmission. Many high-accuracy channel estimation methods have been developed, such as pilot aided linear minimum mean-squared error (MMSE) method, orthogonal matching pursuit method, angle domain channel reconstruction method, and DL-based channel extraction techniques. With the unremitting efforts, the channel estimation accuracy is approaching the Cramer-Rao Bound (CRB). However, existing channel estimation techniques often requires long pilot sequences, especially in massive multi-input multi-output (MIMO) systems, which also leads high costs for providing the CSI feedback. The resources in space, time, and frequency domain are almost exhausted, and new dimensions are urgently needed for channel estimation.

In fact, channel estimation can be regarded as environment sensing, which makes the environment visible. Similar as that for  source transmission, the power of AI and high performance computing devices make it possible to perform the integrated sensing and communications. With further aids of radars, cameras, and sensors, we are able to `see' wireless channels, which does not require the complicated channel estimation process any more. Overall, the generalized semantic communication exploits the semantic domain in both sources and channels to reduce the network traffic significantly without degrading the system performance and lower CSI acquisition cost greatly.

\subsection{Source Semantics}

\begin{figure}[t]
    \centering
    \includegraphics[width=8.8cm]{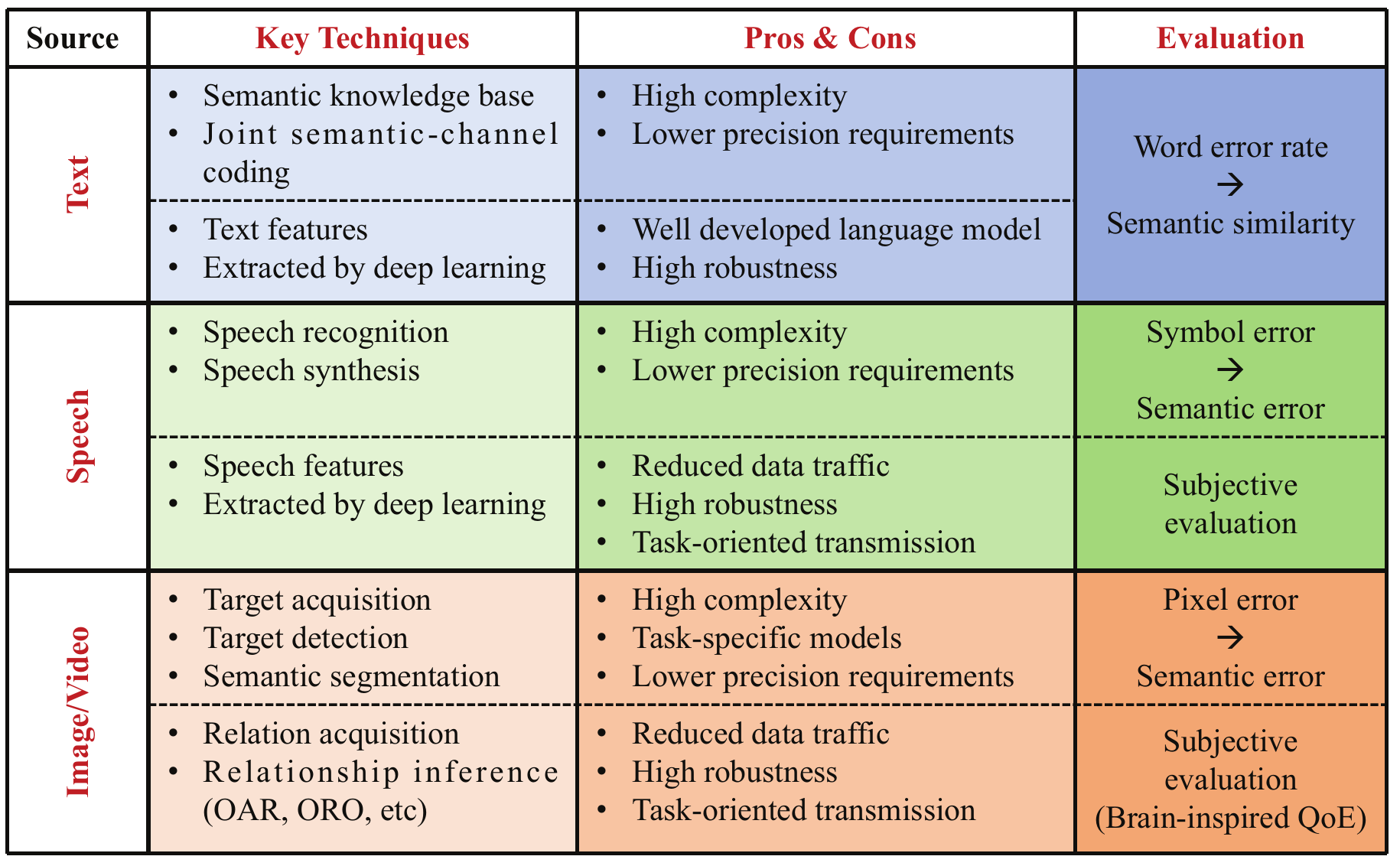}
    \caption{Characteristics of semantics in the source.}
    \label{fig2}
\end{figure}

Semantic communications for source transmission mainly include semantic representation, semantic coding, and semantic transmission. As shown in Fig.~\ref{fig2},  source semantics depend on the source types as discussed subsequently.
\subsubsection{Semantic Communications for Text}
The semantic information of text refers to grammatical information, word meanings, and logical expressions among words, etc. Thanks to the booming of natural language processing (NLP) and DL, the text information can be represented at the semantic level instead of the bit level, which advances the system to extract semantic information from source text. The  goal of semantic communications is to recover these semantic information by minimizing the semantic error. According to a semantic knowledge base shared by the transmitter and the receiver, the semantic representations can be achieved by a DL-enabled joint semantic-channel coding scheme before transmitting over physical channels, which could provide high robustness to both channel impairment and semantic noise.

\textcolor{black}{Different from the Shannon paradigm that utilizes strict bit alignment to measure the system performance, semantic communications require the source information and the recovered information containing the same meaning, which reduces the system accuracy requirement by only focusing semantic information. Moreover, the advancements on NLP facilitates the development of joint semantic-channel coding for text. Nevertheless, the concern of DL-enabled semantic communications for text is the complexity that comes with the DNN training and the computational process significantly increases the transmission latency, which runs counter to the needs of real-time communications.}

In addition to the word-error-rate (WER), new performance metrics to measure the semantic similarity of text are necessary. The bilingual evaluation understudy (BLEU) score is a typical  metric in the machine translation  to measure the quality of the translated text, which has been utilized to measure  semantic errors in  semantic communication systems for text transmission~\cite{9398576}. Moreover, a new  metric, named sentence similarity~\cite{9398576}, has been designed to measure the similarity of two sentences at the semantic level by mapping and then comparing them in the semantic vector space provided by the Bidirectional Encoder Representations from Transformers (BERT) model.

\subsubsection{Semantic Communications for Speech}
\textcolor{black}{The semantic representation of speech is usually more complex than that of text due to the characteristics of speech signals, e.g., the voice of speaker, speech delay, and background noise, etc. Therefore, to exploit the semantic information in speech signals, it is typically processed into a low-dimensional semantic representation. For example, for the speech recognition task, the voice characteristics of speech signals are omitted while the text-related information is leveraged to recognize the correct text sequence. According to the intelligent tasks at the receiver, the corresponding semantic representation of speech signals is extracted by a DL-enabled joint semantic-channel coding scheme~\cite{weng2022deep}.}

\textcolor{black}{Similar to the semantic communications for text, the system performance of semantic communications for speech can be measured at the semantic level. Moreover, in contrast to transmit global speech signal in the typical communication system, the  task-specific speech semantic communication system can be developed by representing input speech sequences into low-dimensional semantics and only transmitting task-related semantic features, which enhance the system robustness as well as significantly reduces the network traffic without task performance degradation. However, the computational burden and DNN training complexity are still the bottlenecks for such a task-specific model.}

\textcolor{black}{The performance evaluation could include objective metrics and subjective assessments. For speech recognition task, the system target is to recover the correct text  from the input speech signals. Character error-rate (CER) and WER are effective to indicate accuracy of the recognized text. For speech synthesis task, unconditional Frechet deep speech distance (FDSD) and unconditional kernel deep speech distance (KDSD) are utilized to measure the similarity of the input  and the synthesized speech. However, the relationship between human perception of speech signals and these objective metrics is not straightforward. Therefore, a dataset has been built to match these performance metric scores with different levels of satisfaction~\cite{weng2022deep}.}

\subsubsection{Semantic Communications for Image and Video}
\textcolor{black}{For the image and video transmission in a semantic manner, the key is to find their structured representations. Different methods have been developed to trace  objectives and their relationships in the image and video so as to represent them in a well-structured way. With visual processing methods, such as target acquisition, target detection, and semantic segmentation, and DL tools, we are able to obtain the relationship between different objectives in the image and video, as well as to perform the further relationship inference. By using such a way to abstract and utilize the semantic information, the source could be compressed to a much smaller size. The structured semantic representation could be  adaptive for severing various intelligent tasks at the receiver in a better way while suffering from the high computational complexity. }

Note that for task-oriented semantic coding or joint semantic-channel coding schemes, the model is  trained specifically for one task. Once the task is changed, the model should be retrained, which causes extra cost for model update. Some general models have been developed to generalize the model so that it is capable of dealing with a group of tasks. Particularly, a unified model has been developed to be applicable for various tasks, which adopts an adaptive structure to allow the receiver exit the model when the current task is executed at a satisfactory level. Such a model is only capable of proceeding a group of tasks. A more general model is more than required to find a good tradeoff between the generalization and system performance.

\textcolor{black}{In comparison to the typical image and video processing methods at the pixel level, such a semantic coding method requires higher computational complexity while training the model and the reconstruction accuracy may not guaranteed when measured at the pixel level. However, the assessment of such an image and video semantic communication system should be redesigned as the typical metrics, i.e., mean-squared error (MSE) and peak signal-to-noise ratio (PSNR), may not reflect the human perception properly. Therefore, quality of experience (QoE) should be considered when measuring the image and video reconstruction quality at the receiver. Particularly, the brain-inspired QoE assessment method provides a good way to find the relationship between the electroencephalogram (EEG) and the subjective perception of  reconstructed images and videos~\cite{HU_GC_2021}. }

\subsection{Channel Environment Semantics}
\begin{figure}[t]
    \centering
    \includegraphics[width=8.8cm]{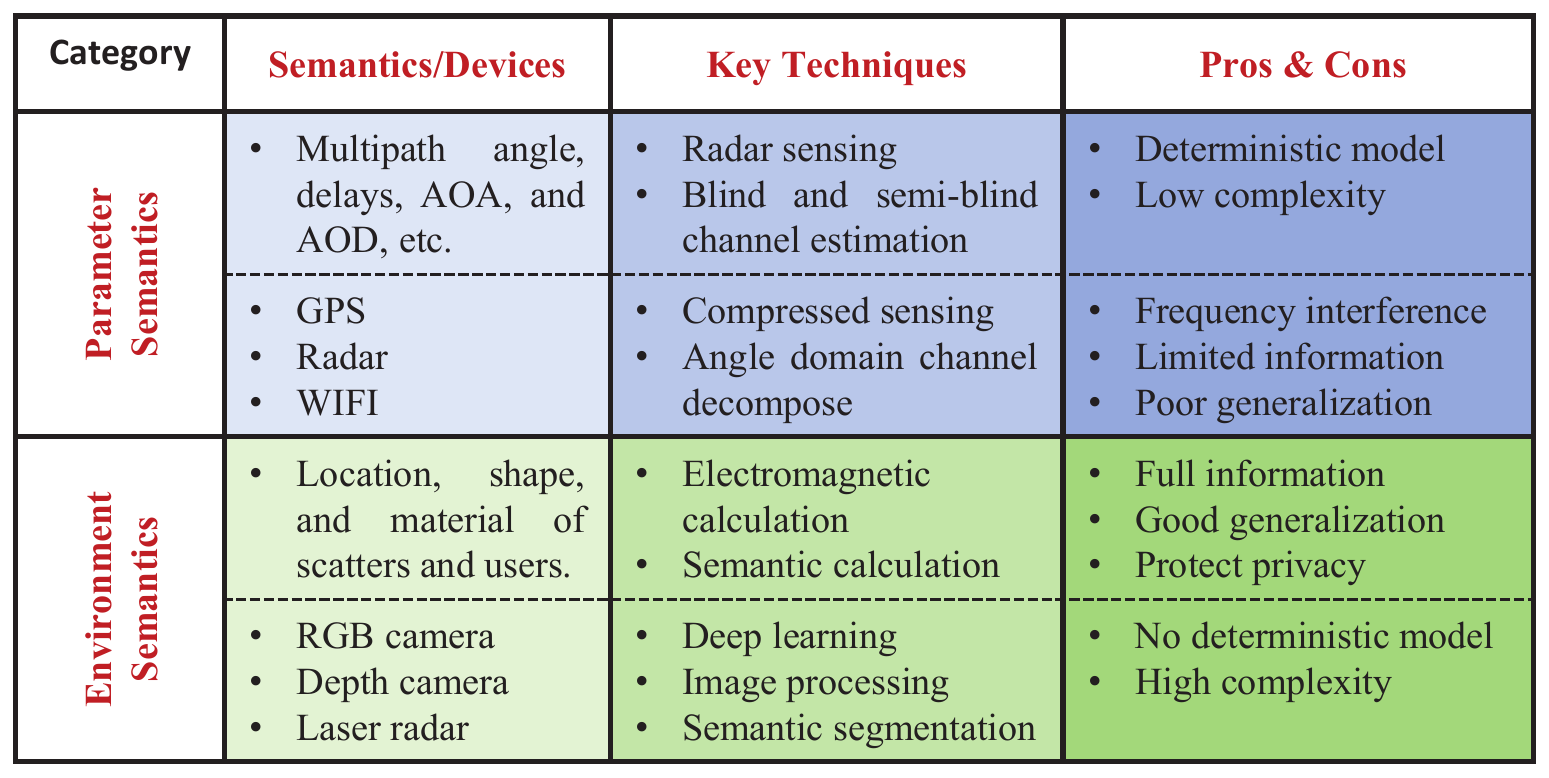}
    \caption{Characteristics of semantics in  channel environment.}
    \label{semantics_in_channel}
\end{figure}

As shown in~Fig.~\ref{semantics_in_channel}, the channel semantics contain two levels: parameter semantics and environment semantics.
\subsubsection{Parameter Semantics}
Parameter semantics refer to parameters that constitute the channel, such as angle of departure (AOD) $\phi_{D}$, angle of arrival (AOA), number of paths, and Doppler frequency offset.
Parameter semantics can be obtained by sensing sensors, such as radar, GPS, and WiFi. In \cite{huang2021mimo}, a radar aided channel reconstruction scheme has been proposed for the \textcolor{black}{multiuser MIMO vehicle to everything (V2X)} communication, where the radar is used for AOD/AOAs estimation while \textcolor{black}{DL-based} algorithm is adopted for channel gain prediction.
\textcolor{black}{Another way to calculate parameter semantics is utilizing signal processing algorithm, such as compressed sensing and angle domain channel decompose techniques, to estimate  parameters from partial CSI}.
In \cite{fan2017angle}, the AOD of a channel is estimated with antenna array theory according to the partial channel matrix. Once the parameter semantics are obtained, the channel can be quickly reconstructed according to the channel model.
Under the guidance of the deterministic model, the channel reconstruction using parameter semantics only requires low computational complexity.
The advantages of parameter semantics are low storage space occupation, low computational complexity and low delay.
However, due to limited amount of information acquired by sensors, obtaining complete parameter semantics requires a large number of sensors.
In order to avoid mutual interference between radar and communication signals, the frequency band used by the radar should be different from that for communications.
In addition, the parameter semantics of different users vary with the frequency although the communication environment is the same.

\subsubsection{Environment Semantics}
Environment semantics refer to  semantic information of  environment images, e.g., the layout, shape, and category of the objects in the images.
%the segmentation results of images and the categories of objects in the environment.
Environment semantics can be captured through RGB cameras and semantic segmentation techniques.
The location and categories of various scatterers can be obtained by semantic segmentation techniques.
With the electromagnetic computing theory, parameter semantics can be calculated from environment semantics, and then the channel matrix can be reconstructed.

Another tool for channel reconstruction is DNN, an efficient means for fitting complex mappings.
We can feed the environment semantics into a DNN, then the DNN generates the reconstructed channel through a series of feedforward operations.
More significantly, the goal of acquiring  CSI benefits the channel-related downstream tasks.
Since there is a mapping relationship between environment semantics and channels, there is also a mapping relationship between environment semantics and the channel-related tasks.
In \cite{9814491}, semantic information is used for cooperative object identification in a multiuser communication system.
Compared with directly using RGB images, utilizing environment semantics to calculate the channel matrix could protect users' privacy as well as save transmission resources. %\cite{zhang2022multi}
This is because that only the categories and layout information of users and scatterers are preserved in environment semantics.
\textcolor{black}{It is noted that  environment semantics are frequency-independent.}
%it shows good frequency generalization.
However, due to lack of deterministic models, the complexity of acquiring environment semantics is extremely high, which consumes huge computing resources.
Although adopting DNN reduces huge computational complexity compared to electromagnetic computing methods, the feedforward operations still require a large number of computing units.
Hence, devices using environment semantics are usually equipped with efficient graphics processing units (GPUs).

%Based on the above discussion, there is a mapping relationship between environment semantics and channel-related tasks.

\subsection{Correlation between Source Semantics and Channel Semantics}
Note that some techniques for processing source semantics could also be used for channel semantics, especially for environment semantics. For example, the deep learning enabled image processing and semantic segmentation techniques are useful for both source semantics and environment semantics processing. Meanwhile, by considering channel semantics, the channel environment could be estimated in a much efficient way. Based on the estimated channel environment, the abstraction of source semantics could be adjusted to fit the current channel condition. For example, if the channel condition is poor, only the most important semantics will be abstracted from the source so as to minimize the size of the transmitted data.

\section{Deep Learning enabled End-to-End Semantic Communications}
Based on the generalized semantic communication system developed in Section II, we first introduce the DL-enabled semantic communication framework in this section. Subsequently, a case study is provided for the speech semantic communication system.

\subsection{\textcolor{black}{Deep Learning Enabled Semantic Communication  Framework}}
\begin{figure}[t]
    \centering
    \includegraphics[width=8.8cm]{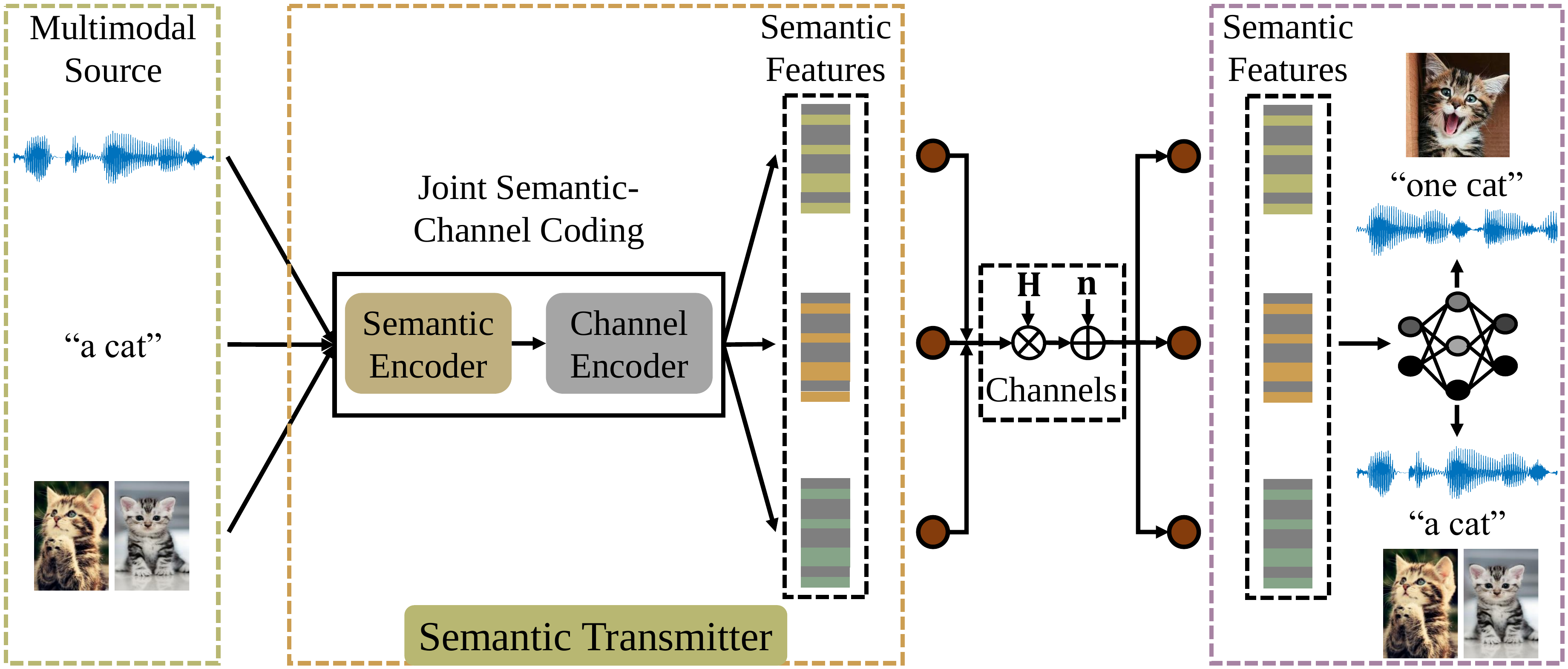}
    \caption{Schematic of the DL-enabled end-to-end semantic communication system.}
    \label{fig4}
\end{figure}

The ultimate motivation of semantic communications is to represent the source information at the semantic level to serve the tasks at the receiver while omits the task-irrelevant information~\cite{qin2021semantic}. By doing so, the bandwidth consumption and the transmission latency can be reduced significantly. A DL-enabled semantic communication system for multimodal data transmission is shown in Fig.~\ref{fig4}. From the figure, a neural network extracts semantic features from the source, relevant to  tasks at the receiver, which could be either source reconstruction or intelligent task execution. Such a framework could be used to serve the transmission of text, speech, images, and videos, which is detailed as below.

\subsubsection{Text}
\textcolor{black}{Based on the DL-enabled semantic communication framework  in Fig.~\ref{fig4}, Xie~\emph{et al.}~\cite{9398576} designed a transformer-powered semantic communication system for text transmission, named DeepSC, by minimizing the semantic difference between the transmitted sentence and the received sentence as well as maximizing the transmission rate. Subsequently, a variant of DeepSC has been proposed in~\cite{jiang2021deep} to further reduce the transmission error by leveraging the hybrid automatic repeat request  to improve the reliability of semantic transmission. More recently, a method to quantify the semantic noise and its affects to the semantic communication system has been found.}

\subsubsection{Speech}
\textcolor{black}{The complexity of semantic representation for speech increases the difficulty to extract semantic features from input speech. Based on DeepSC, Weng~\emph{et al.}~\cite{weng2022deep} proposed a  semantic communication system for speech recognition and speech synthesis, named DeepSC-ST. Particularly, the text-related semantic features are extracted from the input speech and transmitted over physical channel, which serves to recognize the text information and synthesis the speech at the receiver. Inspired by DeepSC-ST, a semantic communication system to fulfill the speech recognition task at the sub-word level has been developed~\cite{han2022semantic}. Moreover, to achieve high accuracy audio transmission with a small amount of data,  federated learning is utilized to allow multiple devices contribute to the semantic information abstraction~\cite{9685654}.}

%Hyowoon~\emph{et al.} design a novel stochastic model of semantic-native communication (SNC) for generic tasks~\cite{seo2021semantics}, in which the semantics are utilized to represent an entity and transmitted to a target listener.
\subsubsection{Image and Video}
\textcolor{black}{Due to the thriving of computer vision, the semantic communications for image have witnessed a rapid development and tackled many challenges beyond human limits. Particularly, based on DeepSC, a semantic communication system for image retrieval has been developed to identify the most similar images stored at the receiver in comparison to the request image~\cite{xie2021task}. Following the  point-to-point semantic communications for transmitting data  with different modalities, Xie~\emph{et al.}~\cite{xie2021task} further extended it to a multi-user case. Moreover, an adaptive semantic coding scheme has been developed to abstract semantic features by considering the rate-semantic-perceptual relationship in~\cite{huang2022towards}. Meanwhile, a semantic communication system has been further developed to support video conference~\cite{jiang2022wireless}.}

%\textbf{In addition, Shao~\emph{et al.}~\cite{9606667} design a semantic communication system for image classification by leveraging a variational information bottleneck (VIB) framework to resolve the difficulty of mutual information computation.}
\subsection{Case Study: Semantic Communications for Speech}
The system model of semantic communications for speech recognition and speech synthesis was developed in~\cite{weng2022deep}, named DeepSC-ST. The input speech is converted into the low-dimensional text-related semantic representation to be transmitted over physical channels, which is achieved by leveraging a joint semantic-channel coding scheme at the transmitter. The receiver includes the channel decoder and the feature decoder to recover the text-related semantic representation and recognize the text information as close to the correct text as possible, respectively. The speech synthesis module processes and converts the output of the feature decoder into the speech sample sequence, which dedicates to reconstruct the clear input speech.

\begin{figure*}[t]
\begin{minipage}[t]{0.48\linewidth}
\centering
\includegraphics[width=0.95\textwidth]{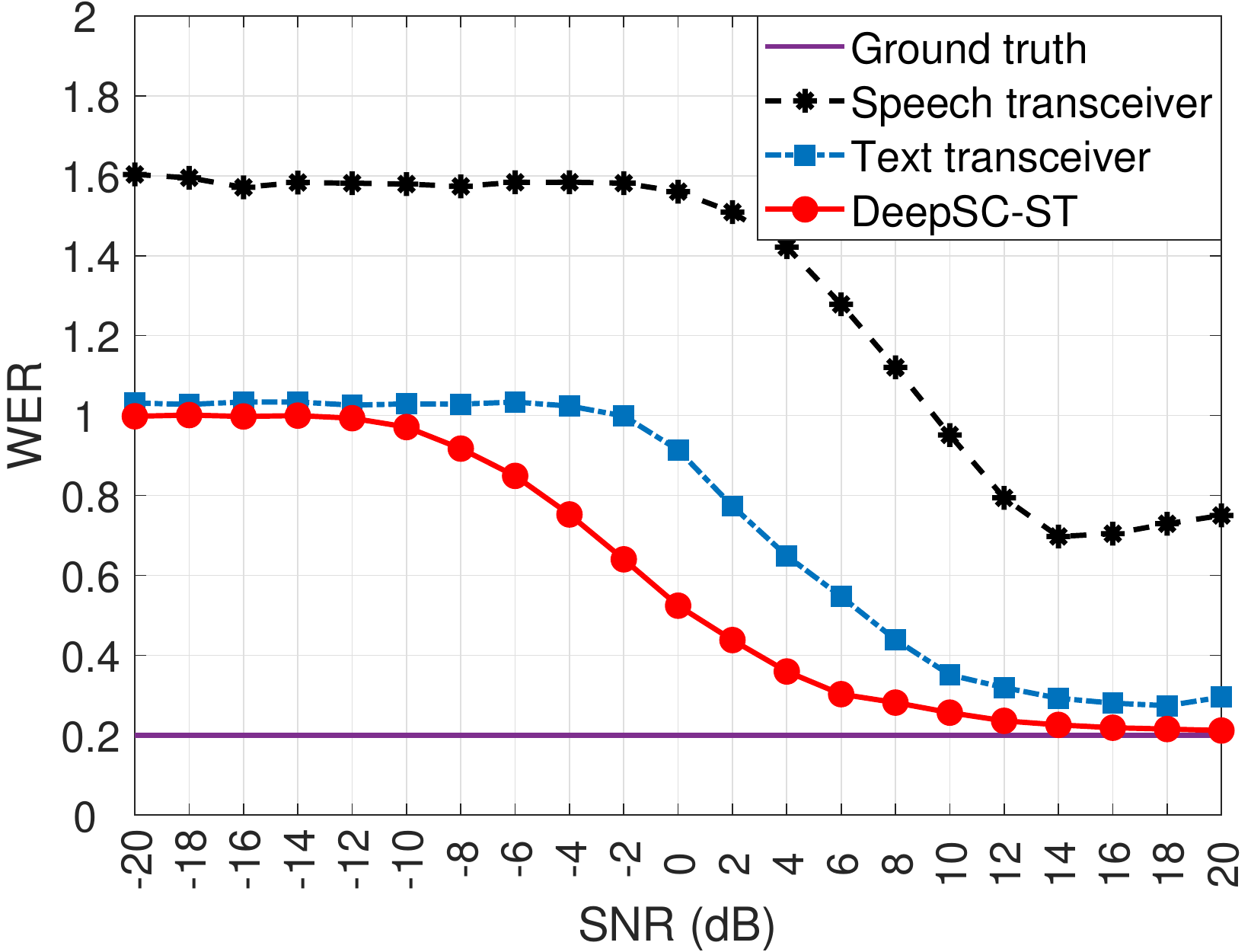}
\subcaption{WER score versus SNR}
\label{WER}
\end{minipage}
\begin{minipage}[t]{0.48\linewidth}
\centering
\includegraphics[width=0.95\textwidth]{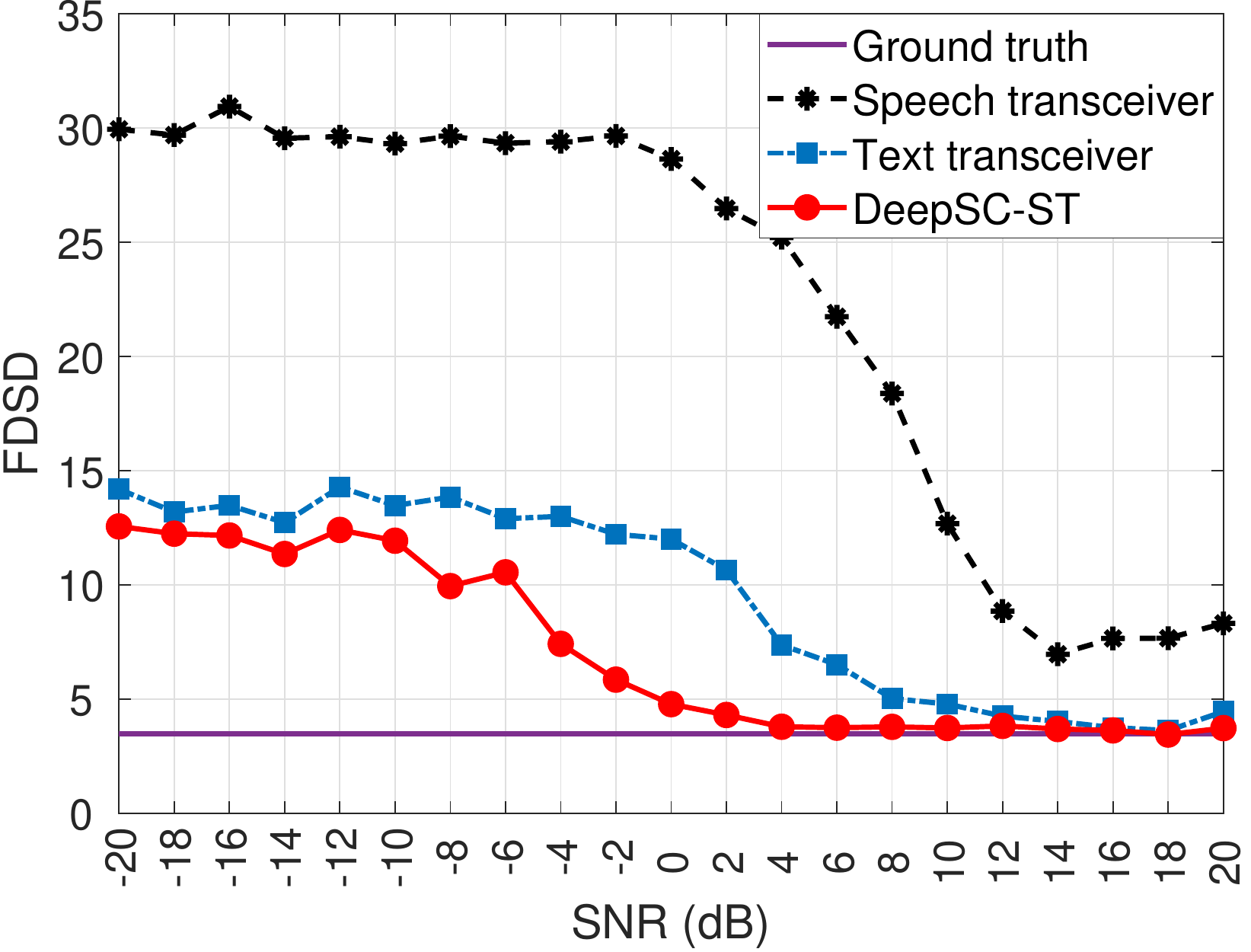}
\subcaption{FDSD score versus SNR}
\label{FDSD}
\end{minipage}
\caption{Performance of DeepSC-ST for speech transmission~\cite{weng2022deep}.}
\label{FDSD result}
\end{figure*}

\textcolor{black}{To verify the superiority of DeepSC-ST, we provide the following two benchmarks. The detailed simulation settings of DeepSC-ST could be found in~\cite{weng2022deep}.}
\textcolor{black}{\begin{itemize}
\item \textbf{Benchmark 1}: we adopts the  convectional communication system with adaptive multi-rate wideband (AMR-WB) coding and polar coding to transmit speech signals, named speech transceiver as shown in Fig.~\ref{FDSD result}. The speech input is transmitted and recovered at the receiver, then the text transcription is obtained from the recovered speech by a speech recognition model, called Deep Speech 2.
\item \textbf{Benchmark 2}: This benchmark is named text transceiver as shown in Fig.~\ref{FDSD result}. Particularly, the speech input is converted into text by Deep Speech 2 before feeding into a conventional communication system with Huffman coding and polar coding  to transmit text. In addition, the recovered text sequence is passed through a speech synthesis model, Tacotron 2, to reconstruct the speech.
\end{itemize}}

\textcolor{black}{Fig.~\ref{WER} compares the WER of the DeepSC-ST and two benchmarks under Rayleigh channels, where the ground truth is the result obtained by directly feeding the speech sample sequence into the Deep Speech 2 model without transmitting them through wireless channels. From the figure,  DeepSC-ST achieves lower WER  than the two benchmarks and performs steadily with SNR$>$0 dB, which verifies the superiority of DeepSC-ST. Moreover,  Fig.~\ref{FDSD} compares the  FDSD scores, where the ground truth is that computed by directly feeding the plain text sequence into the speech synthesis model. From the figure,  DeepSC-ST significantly outperforms  the two benchmarks in terms of FDSD scores, which proves the superiority of DeepSC-ST, especially in the low SNR regime.}

%\textcolor{blue}{According to the DeepSC-ST, a software demonstration with user interface has been developed as a proof-of-concept, as shown in Fig. \ref{DeepSC-ST UI}. In the figure, the real-time speech input is allowed and the communication scenarios with three fading channels under arbitrary SNR values are provided.}
%%%%%%%%%%     Figure: FDSD result     %%%%%%%%%%%%

%\begin{figure}[tbp]
%\centering
%\includegraphics[width=0.4\textwidth]{figs/DeepSC-ST/DeepSC-ST_demo.pdf}
%\caption{Software demonstration of the developed DeepSC-ST.}
%\label{DeepSC-ST UI}
%\end{figure}

\section{Environment Semantics Aided Communications}
This section first introduces the environment semantics aided communication paradigm. Then a case study is provided for the environment semantics aided downlink precoding.

\subsection{Environment Semantics Aided Architecture}
Environment images have been utilized to assist various channel related downstream tasks in some lasted vision aided studies.
However, directly using the environment images to aid wireless communication would violate users' privacy.
The environment semantics aided communications (ESAC) extract semantic information from environment images.
Such semantic information eliminates the surfaces of scatters and preserves only the category and the distribution information, which naturally protects user privacy.

%\begin{figure*}[tbp]
%\centering
%\includegraphics[width=0.8\textwidth]{figs/ESAC.pdf}
%\caption{The structure of environment semantics aided communication.}
%\label{ESAC}
%\end{figure*}

The goal of ESAC is to extract channel semantics from sensors' data, and then use them to predict communication parameter, such as beam index, beamforming vector, blockage state, channel covariance matrix, etc.
%Suppose the parameter to be estimated is $Y$.
Fig. \ref{ESAC} presents the paradigm of environment semantics aided communication.
The ESAC contains four modules: environment semantics extraction module, feature selection module, task-oriented encoder module, and decision module.

\subsubsection{Environment Semantics Extraction Module}
The sensors' data is a set of multimodal observations, including the environment images.
Semantic features are extracted from multimodal data as input to the ESAC.
For example, features from environment images can be extracted by semantic segmentation network.
%Denote the set of all extracted features is $\mathcal{X}_{f}=\{X_{f,1}, X_{f,2}, \cdots, X_{f,i}, \cdots\}$.

\subsubsection{Feature Selection Module}
In fact, not all features are beneficial for communication tasks. Redundant features will increase the complexity of the network or even make the performance worse.
Specifically, the environment semantics contain the class and distribution information of all scatterers in the environment.
However, the features of some objects have little or no effect on the electromagnetic propagation path among the BS and the user.
Moreover, although some features have great influences on the wireless channel, there may be strong correlations between these features. Reserving one of them is enough.
Hence, feature selection is needed to select the optimal set of features, which is the minimum set of features required for the channel related task.
%Suppose the parameter to be estimated is $Y$, which can be
Typical feature selection algorithm includes the wrapper methods, the filter methods, and the embedded methods.
Wrapper methods is based on a specific machine learning algorithm. It follows a greedy search approach by evaluating all the possible combinations of features against the evaluation criterion.
In filter methods, features are selected on the basis of their scores in various statistical tests for their correlation with the outcome variable.
Embedded methods combine the advantageous aspects of both Wrapper and Filter methods.
In embedded methods, feature selection is embedded in the learning or the model building phase. Feature selection is done by observing each iteration of model training phase.
%By using the feature selection algorithm, the most effective feature set is $\mathcal{X}_{s}=\{X_{s,1}, X_{s,2}, \cdots, X_{s,i}, \cdots\}$, where $X_{s,i}$ is the $i$-th selected feature.
\subsubsection{Task-Oriented Encoder Module}
Task-oriented encoder module aims to save the system overhead by encoding the selected features into a compressed one. The task-oriented encoder compress the features as much as possible while loss the key information as little as possible.
In other words, the accuracy of predicting parameter from encoded features and from selected features should be as close as possible.
%Denote the mutual information between X and Y as $I\left(X;Y\right)$.
From a mathematical point of view, the role of task-oriented encoder module is to fit the mapping function such that the number of bits required to store the encoded features is as small as possible while the mutual information between the encoded features and the parameter is as close as possible to that between the selected features and the parameter.
A DNN can be used to realize the mapping function by techniques, such as network pruning, regularization, and model quantization.
\subsubsection{Decision Module}
Decision module predicts the communication parameter from the encoded features.
It is usually implemented by neural networks of different structures according to the output parameter.
For example, for the channel estimation problem, which is a regulation problem, the output parameter is the channel matrix.
The FCN is often utilized as the relization network and the NMSE or MSE function is adopted as the loss function.
For the analogy beam index prediction problem, which is a classification problem, the output parameter is the predicted beam index.
The commonly used networks are FCN and CNN.
The activation function is the softmax function, and the loss function is the cross entropy.

%\begin{figure*}[tbp]
%\begin{minipage}[t]{0.7\linewidth}
%\centering
%\includegraphics[width=1\textwidth]{figs/ESAC.eps}
%\subcaption{The structure of environment semantics aided communication.}
%\label{ESAC}
%\end{minipage}
%\begin{minipage}[t]{0.3\linewidth}
%\centering
%\includegraphics[width=0.95\textwidth]{figs/precoding.eps}
%\subcaption{The average achievable rate of multiuser downlink precoding, SNR = $25$ dB.}
%\label{precoding}
%\end{minipage} %% \end{minipage} put in this line not in the last line
%\caption{The structure of environment aided semantics and the simulation result of the case study.}
%\label{ESAC_and_case}
%\end{figure*}

\begin{figure*}[t]
\begin{minipage}[t]{0.66\linewidth}
\centering
\includegraphics[width=1\textwidth]{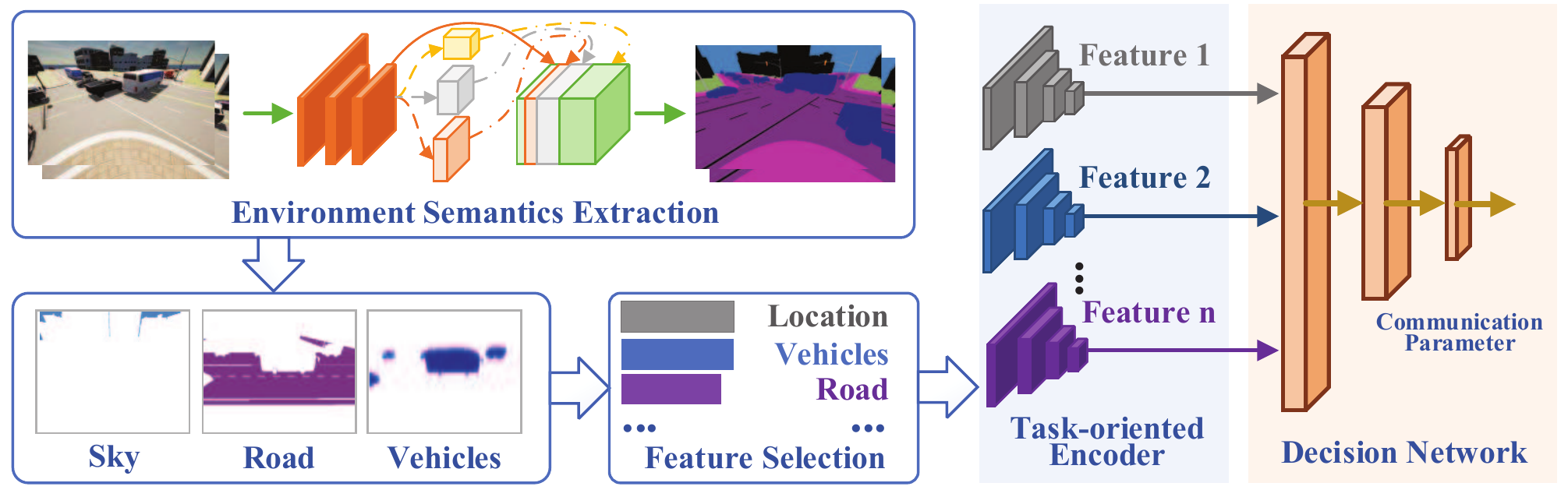}
\subcaption{The structure of environment semantics aided communication.}
\label{ESAC}
\end{minipage}
\begin{minipage}[t]{0.33\linewidth}
\centering
\includegraphics[width=0.9\textwidth]{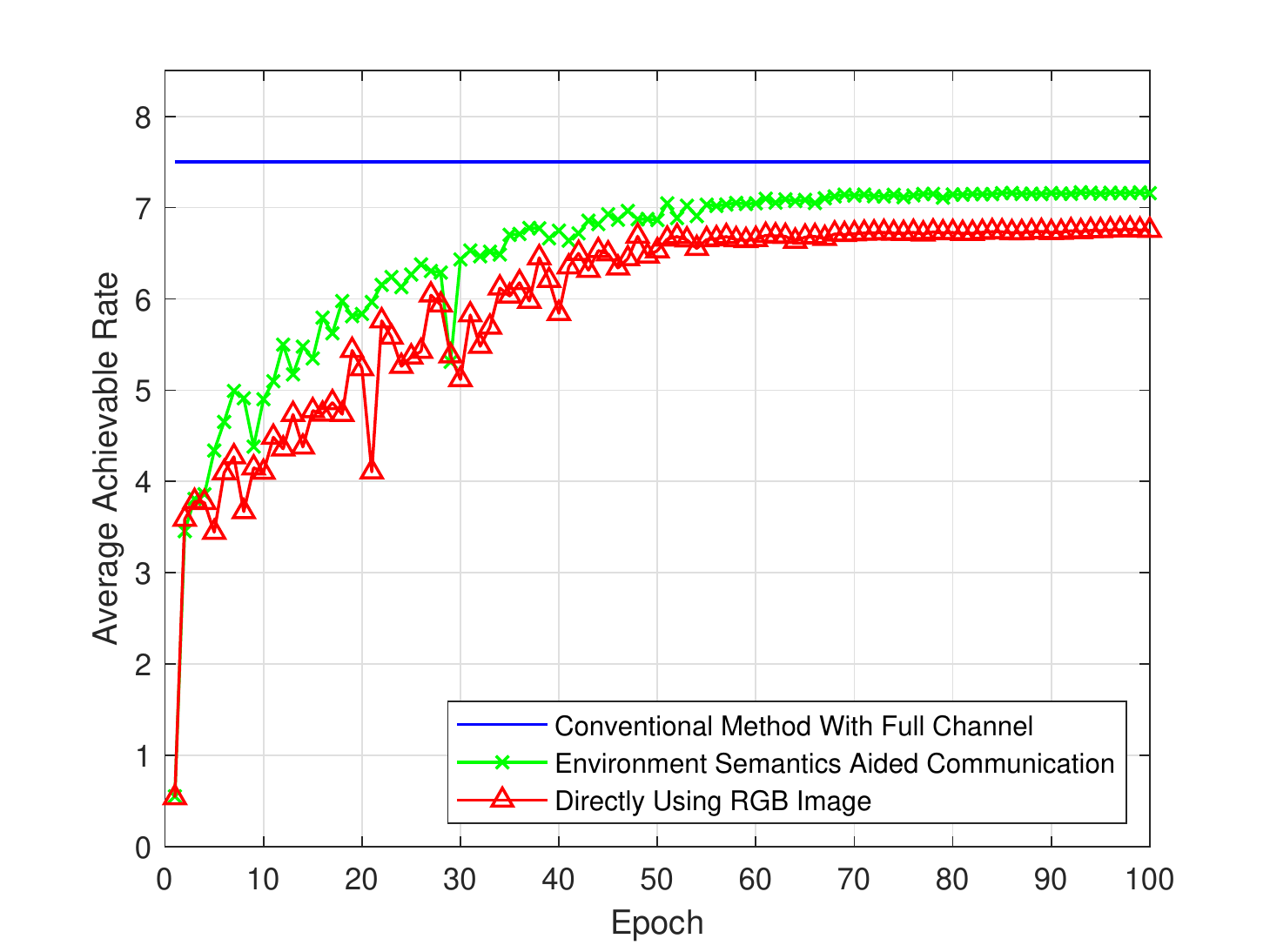}
\subcaption{Rate of ESAC versus epoch.}
\label{precoding}
\end{minipage} %% \end{minipage} put in this line not in the last line

\caption{The paradigm and performance of environment semantics aided communications.}
\label{ESAC_and_case}
\end{figure*}

\subsection{Case Study: Environment Semantics Aided MmWave Precoding}
Consider a multi-user mmwave massive MIMO system where each channel is equipped with one RF-chain.
In order to resist the attenuation of mmwaves and interference between users, the BS needs to perform downlink precoding.
In traditional ways, the BS sends different pilots to different users, then the users perform channel estimation at the receiver end, and then the users feed back the downlink channel to the BS.
Finally, the BS optimizes the precoding matrix according to the users' channels.
However, these methods consume huge training and feedback overhead.
It has been found that mmWave channels usually have a sparse structure associated with the parameters, such as pass loss, AOA, AOD, and delays, etc.
These parameters can be regarded as the parameter semantics.
From an intrinsical perspective, channel is determined by the key scatters in the propagation environment.
Hence, environment semantics are suitable for implementing the multiuser downlink precoding, which is a channel related task.
The goal of this study is to predict the users' downlink precoding matrices directly from the environment semantics.

Suppose that the BS is equipped with multiple antennas and the users are equipped with single antenna.
A camera installed on the BS can capture RGB images to assist the BS on channel related task.
The image is processed by semantic segmentation technology to obtain the environment semantics on the one hand and protect the user privacy on the other hand.
The semantic information includes sky, vehicles, roads, traffic lights, buildings, and lawn.
Using the wrapper method, the semantic features are sorted by importance on the multiuser precoding task.
Then an ODE-inspired network based decision module is designed to predict the multiuser downlink precoding matrix.
Moreover, a user identification network is proposed for user matching by utilizing miltimodal information from the sensing devices.
To increase the system rate, the loss function is the inverse of the average achievable rate of all users in the system.
%The network is extended for any number of users by designing an equivalent loss function.
The achievable rate curves of ESAC network and the network directly using RGB images is shown in the Fig. \ref{precoding}.
From the figure using environment can achieve higher system rate than using images directly.
Moreover, at SNR=$25$ dB, performance of the ESAC can approach the traditional BD algorithm while the training and feedback overhead is reduced to `0'.

\section{Challenges and Conclusions}
This article proposes a framework for generalized semantic communications to fully utilize semantic information in multimodal source data and wireless channel environment. Though various techniques have been developed as detailed in this article to support the generalized semantic communication system, the following challenges should be addressed:
\begin{enumerate}
    \item \textbf{Semantic information transmission techniques}: Most works on semantic communications focus on semantic representation, semantic coding, and its joint design with channel coding. New transmission techniques for semantic features are yet to develop. The existing transmission techniques map bit sequences into symbols without considering the importance of information behind them. To support semantic feature transmission, we need to design novel schemes on semantic-aware modulation and multiple access to combat semantic noise and channel impairment to maximize the transmission efficiency.

    \item \textbf{Unified standard dataset:} Collection of the dataset for environment semantic aided communication is complicated.
    Although there have been specialized datasets for semantic segmentation in the field of image processing, such as Pascal VOC and ADE20K, linking them with wireless communication channel is a challenging work.
    In \cite{lin2022multi}, the vision aided communication dataset has been generated through the cooperative work of three softwares (SUMO, CARLA, Wireless Insite). However, building such a dataset is complex and time-consuming.A convenient user-oriented dataset is urgently needed to speed up the development of ESAC.

    \item \textbf{Generalizability in the number of users:} Environment   semantics contain the semantic information of all scatters and users in the area. However, in a multiuser communication system, the number of users served by the BS changes dynamically. Therefore, for different numbers of users, we need to train different neural networks to perform channel estimation or channel related tasks, which consumes vast computation and storage resources. Investigating more generalized ESAC structures can be an interesting direction for future work.
    %\item \textbf{Lightweight deployment:} Due to the high dimension of the channel matrix, vast parameters are required to train the corresponding neural network. The deployment of the network consumes huge storage resources. Hence, most vision and semantic assisted communication tasks are performed only on the BS. Due to the limited storage resources of the user, it is almost impossible to deploy large neural networks on the user side. However, implementing semantic communication and distributed semantic sharing on the user side has great significance for user scheduling, power allocation, outage prediction, and proactive handoff. Therefore, we need to further study the lightweight environment semantics aided communication structure suitable for deployment on the user side.
    \item\textbf{Generalizability in channel semantics:}
    The generalizability of the channel semantics is guaranteed by the unified feature extraction and feature selection criteria.The sensing data from different sensors has different forms and dimensions. Through feature extraction and feature selection steps, different types of raw data can be  converted into standard features, such as the distribution, categories, and velocities of scatterers. However, there are scenes where the number of sensors is insufficient or sometimes the sensors are malfunctioning. How to utilize the insufficient information to compute channel semantics remains to be studied.
\end{enumerate}

\section{Acknowledgments}
This work was supported in part by the National Natural Science Foundation of China (NSFC Nos 62293484 and 61925105) and in part by Tsinghua University-China Mobile Communications Group Co.,Ltd.  Joint Institute.

\bibliographystyle{IEEEtran}
\bibliography{References}

\section*{Biographies}
\noindent Zhijin Qin (SM'21) is currently an Associate Professor with the Department of Electronic Engineering, Tsinghua University, Beijing, China. Her research interests include semantic communications and sparse signal processing. She served as a Guest Editor and IEEE JSAC special issue on semantic communications an Associate Editor of IEEE Transactions on Communications. She has received several awards including 2017 IEEE GLOBECOM Best Paper Award, 2018 IEEE Signal Processing Society Young Author Best Paper Award, 2021 IEEE Communications Society SPCC Early Achievement Award, and 2022 IEEE Communications Society Fred W. Ellersick Prize.
\\

\noindent Feifei Gao (M¡¯09-SM¡¯14-F¡¯20) is currently an Associate Professor with the Department of Automation, Tsinghua University, Beijing, China. His research interests include signal processing for communications, array signal processing, and artificial intelligence assisted communications. He has authored/coauthored more than 300 refereed IEEE papers that are cited more than 15000 times in Google Scholar. Prof. Gao has served as an Editor of IEEE Transactions on Communications, IEEE Transactions on Wireless Communications, IEEE Journal of Selected Topics in Signal Processing (Lead Guest Editor), IEEE Transactions on Cognitive Communications and Networking, IEEE Signal Processing Letters (Senior Editor), and IEEE Communications Letters (Area Editor).
\\

\noindent Bo Lin received B.S. degree in Electronic Engineering from Xidian University, China, in 2020. He is currently working towards the Ph.D. degree with the Department of Automation, Tsinghua University, Beijing, China, under the supervision of Prof. F. Gao. His research interests include massive multiple-input-multiple-output (MIMO), machine learning, and artificial intelligence assisted communications.
\\

\noindent Xiaoming Tao (SM, IEEE) is currently a Professor with the Department of Electronic Engineering, Tsinghua University. Prof. Tao served as a workshop General Co-Chair for IEEE INFOCOM 2015, and the volunteer leadership for IEEE ICIP 2017. She is serving as the Associate Editor of IEEE Transactions on Wireless Communications and several other journals. She is also the recipient of National Science Foundation for Outstanding Youth and many national awards, such as 2017 China Young Women Scientists Award, 2017 First Prize of Wu Wen Jun AI Science and Technology Award, and 2016 National Award for Technological Invention Progress.
\\

\noindent Guangyi Liu is the Chief Scientist of 6G in China Mobile Communication Corporation (CMCC), Co-Chair of the 6G Alliance of Network AI (6GANA), Vice-Chair of THz industry alliance in China, Vice-Chair of the wireless technology working group of IMT2030 (6G) promotion group. He is leading the 6G R\&D of CMCC since 2018. He has acted as spectrum working group chair and project coordinator of LTE evolution and 5G eMBB in Global TD-LTE Initiative (GTI) from 2013 to 2020 and led the industrialization and globalization of TD-LTE evolution and 5G eMBB.
\\

\noindent Chengkang Pan received his Ph. D. degree from the PLA University of Science and technology in 2008. Currently, he is a senior research member with the Future Mobile technology lab at China Mobile Research Institute. His research interests include 5G and 6G wireless transmission, integrated sensing and communication, semantic communication, and quantum computing for 6G.

\end{document}